\documentclass{aa}  

\usepackage{graphicx}
\usepackage{natbib}
\usepackage{txfonts}
\usepackage[colorlinks=true, citecolor=blue, linkcolor=blue]{hyperref}

\usepackage{color}
\usepackage[dvipsnames]{xcolor}

\definecolor{darkgreen}{rgb}{0,0.5,0}

\begin{document}

   \title{On the periodicity of linear and nonlinear oscillatory reconnection}


   \author{J.O. Thurgood
          \inst{1,2} 
          \and
          D.I. Pontin\inst{1}
          \and 
          J.A. McLaughlin\inst{2}
          }

   \institute{Division of Mathematics, University of Dundee, Dundee DD1 4HN,
              United Kingdom\\
         \and
              Department of Mathematics, Physics and Electrical Engineering, Northumbria University, Newcastle upon Tyne NE1 1ST,
             United Kingdom\\
                           \email{jonathan.thurgood@northumbria.ac.uk}
             }

   \date{Received  3 October 2018; accepted March 11 November 2018}

 
  \abstract
   {An injection of energy towards a magnetic null point can drive reversals of current-sheet polarity leading to time-dependent, \textit{oscillatory reconnection} (OR), which may explain periodic phenomena generated when reconnection occurs in the solar atmosphere. However, the details of what controls the period of these current-sheet oscillations in realistic systems is poorly understood, despite being of crucial  importance in assessing whether a specific model of OR can account for observed periodic behaviour. }
{This paper aims to highlight that different types of reconnection reversal are supported about null points, and that these can be distinct from the oscillation in the closed-boundary, linear systems considered by a number of authors in the 1990s. In particular, we explore the features of a nonlinear oscillation local to the null point, and examine the effect of resistivity and perturbation energy on the period, contrasting it to the linear, closed-boundary case.
   }
   {Numerical simulations of the single-fluid, resistive MHD equations are used to investigate the effects of plasma resistivity and perturbation energy upon the resulting OR. }
{It is found that for small perturbations that behave linearly, the inverse Lundquist number dictates the period, provided the perturbation energy (i.e.~the free energy) is small relative to the inverse Lundquist number defined on the boundary, regardless of the broadband structure of the initial perturbation.
    However, when the perturbation energy exceeds the threshold required for `nonlinear' null collapse to occur, a complex oscillation of the magnetic field  is produced which is, at most, only weakly-dependent on the resistivity. The resultant periodicity is instead strongly influenced by the amount of free energy, with more energetic perturbations producing higher-frequency oscillations. 
   }
   {
Crucially, with regards to typical solar-based and astrophysical-based input energies, we demonstrate that the majority far exceed the threshold for nonlinearity to develop. This substantially alters the properties and periodicity of both null collapse and subsequent OR. Therefore, nonlinear regimes of OR should be considered in solar and astrophysical contexts.   
   }

   \keywords{Magnetic reconnection --
				Magnetohydrodynamics (MHD) --
				Sun: magnetic fields --
				Sun: oscillations --
				Waves 
               }

   \maketitle
   
%

\section{Introduction}\label{sec:intro}
	
	As a fundamental mechanism for energy release in plasmas, magnetic reconnection is of undoubted importance in astrophysics, and has been implicated in a number of phenomena including solar and stellar flares, coronal mass ejections, astrophysical jets, and planetary aurorae.
	In a long history of research into reconnection, since the development of the  `classical', steady-state  2D magnetohydrodynamic (MHD) models such as Sweet-Parker and Petschek reconnection,  a myriad of different models and avenues of research have emerged,
	including the study of 3D effects \citep[e.g.][]{2009PhPl...16l2101P,Pontin3169}, of current sheet instability \citep[e.g.][]{2007PhPl...14j0703L,2016PhPl...23j0702C}, and of kinetic effects \citep[e.g.][]{2010RvMP...82..603Y}, in addition to phenomenological modelling of the large-scale effects of reconnection events on solar and stellar atmospheres \citep[e.g.][]{2000mare.book.....P,priest_book, 2017Natur.544..452W}. 
	
	The solar atmosphere is replete with periodic events. 
	This paper principally concerns periodic phenomena that may occur in time-dependent reconnection schemes, specifically so-called {`oscillatory reconnection}' (OR), which could be characterised as an evolving reconnection region where the directionality of the reconnection undergoes reversals (viz. the current sheets periodically reverse polarity). 
	The idea of an inherently periodic reconnection mechanism is an appealing theoretical explanation for a number of observed (quasi)-periodic phenomena in which reconnection has been implicated, including quasi-periodic pulsations (QPPs) in flares \citep[see][for a review]{McLaughlin2018},
	where OR has recently received attention as a possible explanation \citep[e.g.][]{2016SoPh..291.3427K,2016SoPh..291.3385K,2016SoPh..291.3143V,2017A&A...608A.101P,2018MNRAS.475.2842D,2018ApJ...853....1S,2018ApJ...859..154N}.
In order to test this idea, it is crucial to determine whether OR can produce periodicities compatible with observation under solar conditions (under appropriate parameter ranges and appropriate models). 	However, exactly what controls this period is currently poorly understood, and so this paper seeks to contribute towards clarifying this issue.

	Perhaps the oldest time-dependent reconnection model is that of `null collapse' (or `X-point collapse').
	The basic idea is that perturbations should tend to collect in the close vicinity of null points due to a refraction effect \citep[see wave-null interaction studies, e.g.][]{2011SSRv..158..205M,2012A&A...545A...9T,2013A&A...558A.127T},
	 and then participate in an implosive process which may produce high current concentrations on small scales where dissipation can become effective (first proposed by \citealt{dungey53}).
	Before the 1990s, null collapse had already received significant attention \citep[as attested to by the posthumously-published review of][]{syrovatskii1981review} 
as it was thought that the scaling of the current growth during the implosive phase could provide  an avenue to `fast' reconnection, even in highly-conducting plasmas. 
	Though informative with regards to the initial implosive phase of null collapse, these studies relied on similarity solutions which break down at the scale at which diffusion (or other processes) could limit such an implosion, and therefore could not follow the evolution beyond this time (hence, to possible OR).
	Thus, we do not discuss these specific results much further here  \citep[see introductions of][for a recent perspective on the history of null-collapse research]{Thurgood2018a,Thurgood2018b}. 
	
	In the 1990s, a significant amount of research was published that considered null collapse using a variety of distinct approaches that were (in-part) motivated by overcoming some of the questions surrounding the physicality of  the earlier similarity-based solutions, by instead using perturbation-based techniques.
	Notably, 	\citet{1991ApJ...371L..41C} 
\citep[and subsequently, e.g.][]{1992ApJ...393..385C,1992ApJ...399..159H,1993ApJ...405..207C} produced models and solutions which concerned the evolution of a collapsing null due to a perturbation of \emph{fixed} free energy contained within a finite, closed domain, inclusive of the field surrounding any similarity region that may form naturally (dynamically). 
	Such solutions could follow the evolution beyond the initial implosion, and indicated that OR would occur at the null point. Henceforth, we refer to these papers and the general scheme or type of OR they concern as `linear OR'.
		A summary of these linear analyses is given in \citet[][Chapter 7.1]{2000mare.book.....P}. 

	In this {particular scheme} of OR (`linear OR'), the oscillation is essentially a global oscillation of the magnetic field set up as a consequence of MHD waves  repeatedly reflecting between the diffusion region (a small shell around the null of radius proportional to $\eta^{0.5}$) and the closed domain boundary. 
	As a consequence, \citet{1991ApJ...371L..41C} found that the  period associated with such OR is set by the communication time between the boundary and the diffusion radius.
	As such, the periodicity may be expressed in terms of a single variable, namely the (inverse) Lundquist number defined in such a way to encode this communication time (i.e. a Lundquist number based on values of typical speed at the reflecting boundary, length scale to the boundary, and resistivity).
	 They also determined a corresponding decay rate, which was similarly simple. Later it was found by \citet{1992ApJ...393..385C} that this decay rate does not apply in the general broadband oscillations, however, they did find that the period itself was unchanged regardless of the initial condition. 

	Since consideration in the 1990s, the OR phenomenon has been reported and studied in a number of distinct contexts and models.
	 For example, in simulated model solar atmospheres, periodic  current sheet reversals have been noted to occur in response to flux-rope emergence \citep{2009A&A...494..329M,2012ApJ...749...30M} and  p-mode driving  \citep{2017ApJ...837...94T}. 
	 OR has also been studied as a response to externally-originating shock waves converging upon a null \citep{2009A&A...493..227M,2012A&A...548A..98M}, and also in a similar setup extended to Hall MHD by \citet{2012A&A...544A..24T}.
	 Further, in the first such study, we recently confirmed that OR is permitted about three-dimensional (3D) null points, occurring in the (repeating) `spine-fan' mode of reconnection \citep{Thurgood2017}. 
	 	 
	 	 It is clear that none of the oscillations reported in these 21st century papers are compatible with the global oscillation periods in the linear OR models.
	 	 These simulations are rife with nonlinearity, and a number of them claim to have effectively implemented a non-reflecting boundary which would preclude reflection-driven oscillations altogether  by perfect transmission of outgoing waves (e.g. \citealt{Thurgood2017} reported rigorously on how this was achieved).
	 Most of these papers did not explicitly consider what controls the periodicity of the resulting oscillation in their experiment by parameter variation, more commonly just reporting on the OR phenomenologically. 
	 An exception is \citet{2012A&A...548A..98M}, which considered the effect of varying the pulse amplitude in the converging-shock triggered case 
	 (propagating shock fronts act to collapse a null and form the initial current sheet, which then oscillates),
	 and found that the period of the long-term oscillation decreased with increasing amplitude (in the parameter range considered).
	 This is of course in contrast to the linear OR models where the period is set \emph{only} by the Lundquist number, clearly indicating that it is not a formula which applies to all systems exhibiting OR, even as an approximation.  

	Understanding what controls the periodicity of OR in realistic systems is of crucial importance if we are to begin to test these theoretical models of OR against observation (\textit{``What periods can be produced in solar parameter ranges?''}). 
	Evidentially, from the more recent simulations, the formula of \citet{1991ApJ...371L..41C} does not seem appropriate as the `realism' of the models increases.
	 As such, in this paper we have two goals. 
	 The first is to explore why this is the case, determine under what conditions the periodicity of OR departs from that of linear OR, and to characterise the qualitative properties of such a periodicity upon this departure. 
	 As we will see, this is primarily a consequence of the nonlinear properties of null collapse which are absent in the linear OR models. 
	 Crucially, we show that for astrophysical parameters  the perturbation energy threshold for nonlinear evolution is so small that it seems unlikely that any astrophysical periodic reconnection phenomena of interest will involve linear collapses (and thus conform to the linear OR regime).  
	   The second aim is to further our understanding of the more applicable nonlinear regime of OR.
	   Thus, in a parameter study we consider the influence of free energy (perturbation energy) and --- for the first time --- the influence of plasma resistivity on the resulting period.

\section{Numerical setup}\label{sec:setup}
	The simulations involve the numerical solution of the single-fluid, resistive MHD equations using the LareXd code \citep{2001JCoPh.171..151A}.
	Here we outline the simulation setup (initial conditions), with full technical details deferred to the appendicies. 
	All variables in this paper are nondimensionalised, unless units are explicitly stated.
	
	We consider a background magnetic field of the form 
	\begin{equation} \label{eqn:b0}
		\mathbf{B}_{0} = \left[ y, x, 0\right] \quad,
	\end{equation}		
	which is a potential null point, free from electrical currents, and so constitutes a minimum-energy, force-free state. 
	This field is embedded within a plasma we take to be initially at rest ($\mathbf{v=0}$),
	 of uniform density ($\rho=1$) and uniform gas pressure,
	 chosen such that a fixed plasma-$\beta$ defined by the background field $\mathbf{B}_{0}$ at radius $r=1$ may be set, which is taken as $\beta_{0}=10^{-8}$. 
	 We consider an ideal gas with $\gamma=5/3$.
	 Plasma resistivity $\eta$ is also taken as a uniform variable in our parameter study.
	
	To this field we consider a perturbation $\mathbf{B}_1$ (such that the total field $\mathbf{B}=\mathbf{B}_0 +\mathbf{B}_1$) that is of the form 
	\begin{equation} \label{eqn:b1}
		\mathbf{B}_{1} = \mathbf{\nabla}\times\mathbf{A}_{1}\,;\,\mathbf{A}_{1}=\frac{1}{2} \cos \alpha \left(1-x^2\right)\left(1-y^2\right) \hat{\mathbf{z}} \quad ,
	\end{equation}		
	which corresponds to a nearly-uniform current concentration centred about the null which is then tapered down (with associated return current) such that the flux at the boundary is undisturbed. This is essentially the form of the perturbation used in the simulations of \citet{1996ApJ...466..487M}. 
	 We express the initial amplitude of this perturbation through $j_0$ which is the peak initial current density associated with the perturbation (equivalently, we control the available free energy through setting $j_{0}$), which is related to the initial separatrix angle $\alpha$ by $j_{0} = 2\cos\alpha$.
	 As the boundary conditions are closed and line-tied (Appendix \ref{app:boundary}) this initial condition contains all of the free energy available to participate in the resulting evolution. 
	
	Such a perturbation imbalances the Lorentz force about the null, and triggers {`null point collapse'} --- an implosive process where the perturbation energy is focused to increasingly small scales where dissipative processes and magnetic reconnection can become (momentarily) significant, even in extremely-high-conductivity plasma.
	If the perturbation amplitude at this time is sufficiently small (relative to the resistivity),
	the waves and dynamics of the initial implosion proceed in a linear regime 
	{(for this specific initial condition,  we find the nondimensional condition $j_{0}\lesssim2.1\eta$ is required for linear collapse, see Appendix \ref{app:threshold} and Section \ref{sec:dimensions}).}
	Otherwise, a nonlinear implosion characterised by different scaling laws and the production of extremely localised heating, dissipation and  plasma inhomogeneity (including shock structures at the critical time) will occur.
	We have recently discussed  the differences between these initial collapses in some detail \citep[][see also references therein]{Thurgood2018a,Thurgood2018b}. 
	Regardless of linearity, after this initial implosion stalls, OR begins to occur as the system seeks equilibrium.

   \begin{figure}
   \centering
   \includegraphics[width=\columnwidth]{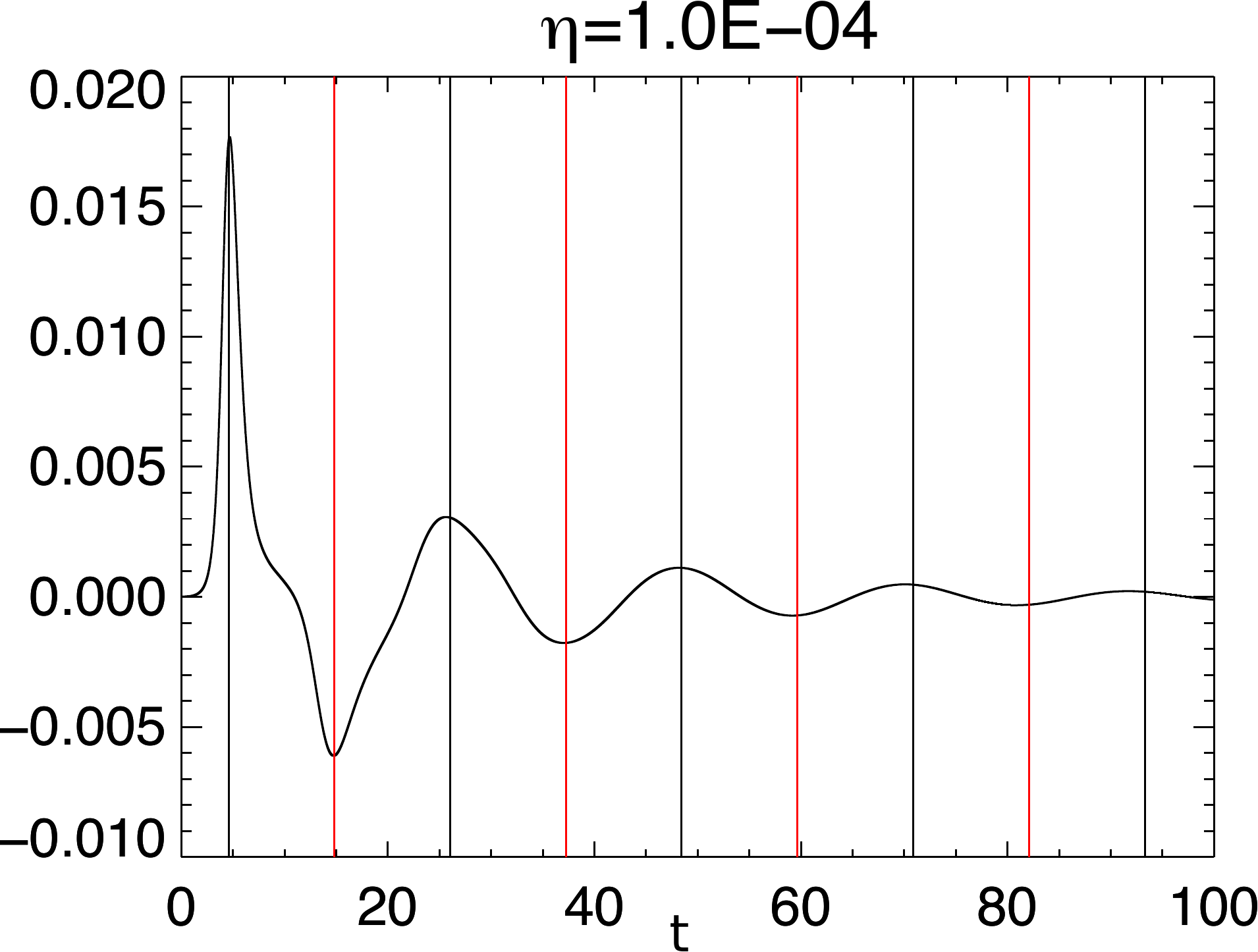}
      \caption{Evolution of the current density $j_{z}$ recorded at the null during the $j_{0}=10^{-5}$, $\eta=10^{-4}$ simulation. The vertical lines indicate the locations of the predicted local extrema according to {{Equation (\ref{eqn:linprediction})}} where $a=0.926$ is determined empirically (see description of Table \ref{tab:linear}).}
         \label{fig:linearexample}
   \end{figure}	
   
      \begin{figure}
   \centering
   \includegraphics[width=\columnwidth]{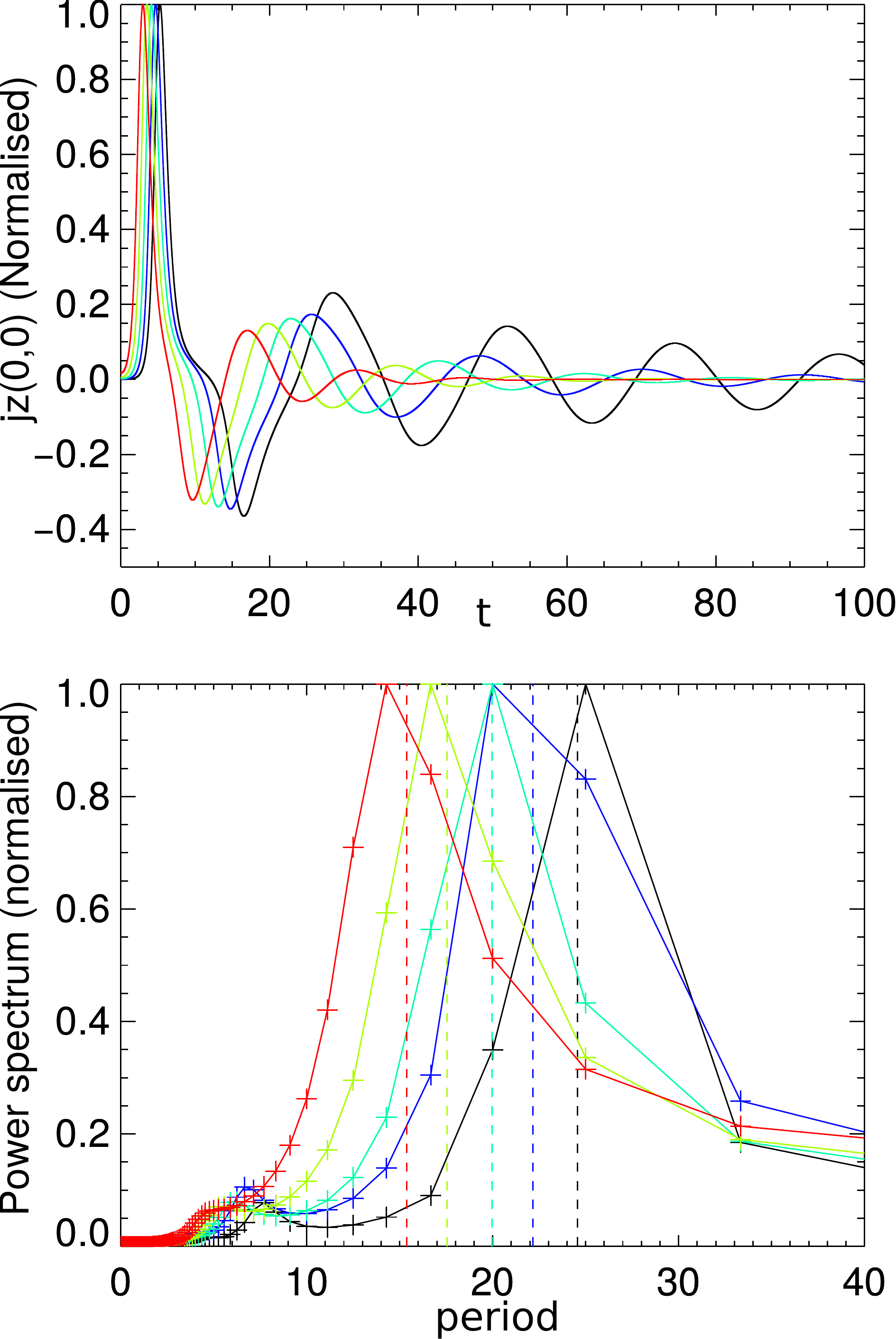}
      \caption{\textit{Top:} Curves of normalised $j_{z}$ recorded at the null during {linear} oscillatory reconnection ($j_{0}=10^{-5}<{\eta}$) for a range of resistivity from $\eta=3\times10^{-5}$  (black) through $\eta=1\times10^{-3}$  (red) as per Table \ref{tab:linear} (following maxima left to right, the curves go from higher to lower resistivity). 
      \textit{Bottom:} Normalised Fourier power spectrum indicating the dominant period measured in the $j_{z}(0,0)$-signals above. The vertical dashed lines indicate the expected period according to Equation {{(\ref{eqn:linprediction})}}, which is consistent with the dominant periodicity bin in all cases. }
         \label{fig:linearsignals}
   \end{figure}	
	
\section{Linear oscillatory reconnection}\label{sec:linear}

We first present the results of simulations of OR that meet the requirements to evolve in the linear regime (to which the results of Craig and co-authors should apply)  in order to both (i) elucidate the key results regarding what controls the periodicity under such conditions, (ii) validate our numerical setup in this better-understood regime before proceeding to examine what happens when we violate the linearity condition by considering higher-energy perturbations in 
Section \ref{sec:NL}, and (iii) provide a point of contrast for these nonlinear results.
We first construct an analytical prediction of the period (Section \ref{sec:linearperiod}), and then compare this to our simulation in Section \ref{sec:linearsims}. 

\subsection{Linear OR: Analytical prediction of the period}\label{sec:linearperiod}

\begin{table*}
\centering
\caption{Predicted and measured periodicities in the linear, closed-boundary simulations (see Section \ref{sec:linearsims} for a list describing the different column headers). In both the case of the predicted periods for $a=1$ and an empirically corrected period for $a\neq1$, the prediction is consistent with the peak spectral power bin, although by comparison to the average distance between local extrema the adjusted prediction appears to have closer agreement to simulation. }\label{tab:linear}
\begin{tabular}{ccccccccc}
\hline
$\eta$     & $j_{0}$ & $t_c$ ({{$a$}}=1) & $t_c$ (measured) & $a$ (empirical) & $P$ ({{$a$}}=1) & $P$ (empirical $a$) & $P$(FFT)        & $P$(extrema) \\
\hline
0.00003 &  $10^{-5}$  & 5.2072              & 5.2824        & 0.928         & 24.829            & 24.571                     & 22.22 - 28.57   & 23.08      \\
0.0001  &  $10^{-5}$  & 4.6052              & 4.6814        & 0.927         & 22.421            & 21.160                     & 18.18 - 22.22 & 21.67      \\
0.0003  &   $10^{-5}$ & 4.0559              & 4.1318        & 0.927         & 20.223            & 19.963                     & 18.18 - 22.22 & 19.43      \\
0.001   &   $10^{-5}$ & 3.4539              & 3.5287        & 0.928         & 17.816            & 17.559                     & 15.39 - 18.18 & 16.86      \\
0.003   &   $10^{-5}$  & 2.9046              & 2.9765        & 0.931         & 15.618            & 15.370                     & 13.33 - 15.39 & 14.03      \\
\hline
0.00003 &  $0.01{\eta}$  & 5.2072              & 5.2832        & 0.927         & 24.829            &  24.571                    & 22.22 - 28.57   & 24.20      \\
0.0001  &   $0.01{\eta}$ & 4.6052              & 4.6815        & 0.927         & 22.421            &   21.160                   & 18.18 - 22.22 & 21.81      \\
0.0003  &  $0.01{\eta}$  & 4.0559              & 4.1319        & 0.927         & 20.223            &   19.963                   & 18.18 - 22.22 & 19.59      \\
0.001   &  $0.01{\eta}$  & 3.4539              & 3.5287        & 0.928         & 17.816            &  17.559                    & 15.39 - 18.18 & 17.02      \\
0.003   & $0.01{\eta}$   & 2.9046              & 2.9764        & 0.931         & 15.618            &  15.370                    & 13.33 - 15.39 & 14.71     \\
\hline
\end{tabular}
\end{table*}

		For a linear disturbance to the null, we expect that the period of a standing or global oscillation is essentially set by the (fastest) communication time between the boundary and the null point. By assuming a cylindrical boundary,
		 recognising that the linear wave speed for the setup simplifies to $c_{A}=r$ (see Equation \ref{eqn:b0}),
		 and solving by separation of variables we may calculate the time taken for a cylindrical mode to traverse from the outer boundary (at $r=1$) to a diffusion-dominated interior layer.
		 By dimensional arguments, this `{diffusion radius}' is  $r_{c} = a\eta^{1/2}$ where $a$ is a constant of order 1 
		 {\citep[see][]{1991ApJ...371L..41C,1992ApJ...393..385C}}
		  Making no assumption on $a$, this communication time is
	\begin{equation}\label{eqn:tcprediction}
	t_{c} = \frac{1}{2} \ln \left( \frac{1}{\eta} \right) + \ln \left( \frac{1}{a} \right) .
	\end{equation}
	This is sometimes referred to as the `{critical time}' in null collapse literature, as it corresponds to the time at which diffusion begins to dominate the dynamics of the initial collapse promoted by our initial condition, and the time at which the maximum current density and reconnection rates will be produced \citep[see][for further details]{Thurgood2018a,Thurgood2018b}. 
	For now, we neglect the time it may take for information to diffuse through this zone, and simply argue that a complete period of a global (standing) cylindrical oscillation of the null must be four times this quantity
	\begin{equation}
	T_{\textrm{wave}} = 2 \ln {\left( \frac{1}{\eta} \right)} +  4 \ln {\left( \frac{1}{a} \right)} .
	\end{equation}	 
If we choose to use the approximation $a\approx1$,  continue neglecting to account for a diffusion time, and recognise that $\eta^{-1}$ is the Lundquist number under our nondimensionalisation (see Appendix \ref{appendixA}) this simplifies to 	
\begin{equation}\label{CM91period}
	T_{\textrm{wave}} = 2 \ln S
\end{equation}
	which is the period found in the literature \citep[see, e.g.][]{1991ApJ...371L..41C}.  
	Thus, our back-of-envelope construction of an expected period is consistent with {bona fide} solutions of the linearised
	 system. 
	In this form (equation  \ref{CM91period}) it is most clear that in this regime of OR the oscillation's period is uniquely determined by the time taken for reflections of radial oscillations propagating between the boundary and the interior diffusive layer. 
	Hence, it depends only on the Lundquist number (i.e.~the {Alfv{\' e}n} speed on the outer boundary, the length scale to the boundary, and the resistivity which sets the scale of the interior boundary) which under this system of nondimensionalisation is characterised by the single variable $\eta$. 

We may attempt to account for the time taken to traverse the diffusive layer to the origin by assuming a diffusive speed of $v_{d}=\eta/r_{c}$, which since $r_{c} = a\eta^{1/2}$ yields 
\begin{equation}
t_{\eta} = \frac{r_c}{\eta/r_c} = a^{2}. 
\end{equation}
We note that, although this is relatively small compared to $T_{wave}$ for astrophysical (large) Lundquist numbers, it is important to account for this sometimes neglected timescale if we wish to compare this formula to full MHD simulation. This is because, due to computational limitations, the simulated explicit resistivity will  be comparatively high in order to allow for the diffusion layer to be properly-resolved on the grid (recall $r_{c}=a\eta^{1/2}$). Thus, our overall prediction for the period of linear OR is 

\begin{equation}\label{eqn:linprediction}
 P = 4t_{c} + 4 t_{\eta} = 2 \ln \left( \frac{1}{\eta} \right) +  4 \ln \left( \frac{1}{a} \right) + 4 a^{2} .
\end{equation}

\subsection{Linear OR: Simulated periodicities}\label{sec:linearsims}

	We consider a set of simulations which essentially exhibit linear OR and compare them to the predictions.
	To recover the linear behaviour throughout it is sufficient that the perturbation energy satisfies
	{$j_{0}\lesssim2.1\eta$}, i.e., the criterion for the initial implosion to stall \emph{before} nonlinear steepening occurs.  
	As such, we consider runs with $\eta$ in the range $\eta = 3\times10^{-5}-3\times10^{-3}$ and initial amplitudes $j_{0}=10^{-5}$ (i.e. a fixed energy that satisfies the condition across the range considered), and also $j_{0}=0.01\eta$ (a variable energy that is fixed in its ratio to nondimensional $\eta$). 
	Note that the simulations themselves are of the full nonlinear equations, rather than a linearised reduction.
	
	In Figure \ref{fig:linearexample}, as an example we show the evolution of the normalised current density at the null point (located at the origin) over the span of $100t_{A}$ for $j_{0}=10^{-5}$ for
	{ $\eta=10^{-4}$}. 
	The signal of $j_{z}(t)$ initially forms a peak associated with the collection of the broadband driver into a small radius (area) during the initial (linear) implosion. 
	Note that our initial condition is not entirely cylindrically symmetric, nor are the boundaries, thus affecting the accumulation of current slightly as per \citet{1992ApJ...393..385C}.
	 After, it assumes a damped,  regularly-sinusoidal	profile where the change in the sign of $j_{z}$ and so the reconnection rate ${\eta}j_{z}(\mathbf{0},t)$ indicates the occurrence of OR. 
  	This signal is typical of all of the simulations of linear OR.

	The measured and expected periods of this signal  in $j_{z}$ for all of the  different (linear amplitude) runs is summarised in Table \ref{tab:linear}. Specifically, it quantifies 

	\begin{itemize}
	\item The critical time $t_{c}$ expected from Equation (\ref{eqn:tcprediction}) under the assumption that $a=1$ for the given value of $\eta$ in each run.
	\item The measured critical time from the simulated data (`$t_{c}$ measured') .
	\item The value of $a$ necessary to match the predicted critical time from Equation (\ref{eqn:tcprediction}) to the data,  henceforth referred to as the `empirically adjusted value of $a$'.
	\item The expected period under the $a=1$ assumption (`$P(a=1)$').
	\item The expected period under the empirically-adjusted value of $a$ (`$P$(empirical $a$)'). 
	\item The measured period according to the periodicity bin corresponding to peak Fourier power calculated from the simulated $j_{z}$ signal at the null (`$P$(FFT)').
	\item The measured period by the average time between local maxima in the simulated $j_{z}$ signal at the null (`$P$(extrema)').
	\end{itemize}	

		 We see excellent agreement of the predicted period (equation \ref{eqn:linprediction} with different choices of $a$) and the measured periods (and also, by visual inspection).
	 	Despite the fact that we use a Cartesian boundary and the fact that the full solution permits broadband oscillation (i.e. can contain contributions from $m\neq0$ modes), the measurements of the period do not significantly depart from this  prediction (this is also in agreement with arguments by  \citealt{1992ApJ...393..385C} who present a linear solution inclusive of higher modes).
	 			 We also note that in both sets of amplitudes in the runs ($j_{0}=10^{-5}$ and $j_{0}=0.01\eta$) that the periods are   as predicted by Equation (\ref{eqn:linprediction}) --- indicating that the period is independent of the wave amplitudes (as would be expected under a linear evolution controlled by fixed background characteristics/wave-speeds). The actual quantitative difference between the corresponding values of, e.g., $t_{c}$, is small, as evident in the tabulated data.

 	Figure \ref{fig:linearsignals} (top panel) also visualises the various signals for the fixed-amplitude $j_{0}=10^{-5}$ and variable $\eta$ set of runs, where the increase of the period with decreasing resistivity is clear (it also shows that higher resistivity increases the damping rate, due to enhanced dissipation, although we do not focus on this aspect of OR further in this paper). 
  	In the bottom panel of 	Figure \ref{fig:linearsignals}, we show normalised Fourier power spectra calculated from these signals, with the vertical-dashed lines indicating the predicted period which corresponds in all cases to the periodicity bin with the dominant power (adjusted for empirical $a$, although the $a=1$ prediction is also consistent with the spectral measurement).

		Overall, the periodicity of OR in a linear regime is well-understood and documented in the works of {\citet{1991ApJ...371L..41C}} and subsequent authors. 
		We have demonstrated in this Section that we can reproduce the key properties of such a system in numerical simulation, `validating' our numerical setup against independent predictions.
		This particular case (linear OR) is subject to a number of assumptions, restrictions, or criteria, perhaps most stringently the linear collapse condition itself. 
		{We discuss the physical meaning of this condition further in Section \ref{sec:dimensions}, and find that it is unlikely to be satisfied in astrophysical plasmas. }
		{Examples of other restrictions on the applicability of OR as described} in this section include perfect reflectivity, and the numerous conditions for collapse to be linear under the action of other limiting mechanisms {such as guide-field back-pressure} are discussed in detail in \citealt{Thurgood2018a,Thurgood2018b}).

\section{Nonlinear oscillatory reconnection}\label{sec:NL}

   \begin{figure*}
   \centering
   \includegraphics[width=13cm]{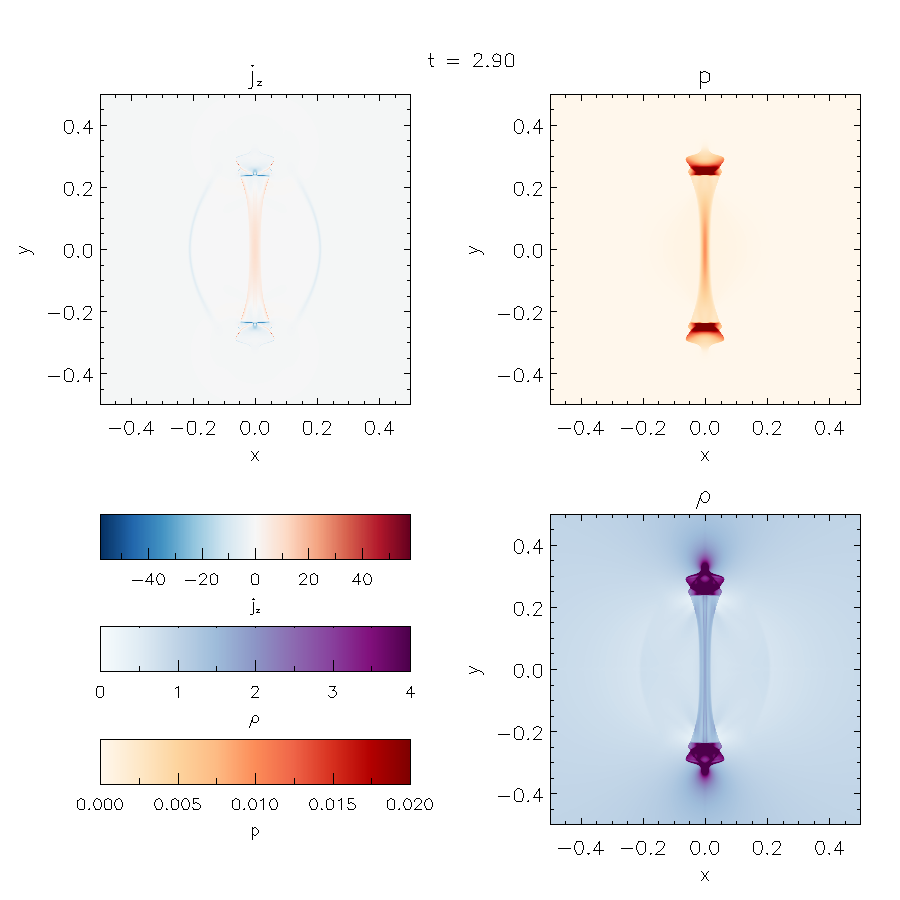}
      \caption{Evolution of  $j_z$, $\rho$, and $p$ about the null point, for the initial conditions $\eta=10^{-4}$, $j_{0}=0.1$ (animated). }
         \label{fig:NLmov}
   \end{figure*}

   \begin{figure}
   \centering
   \includegraphics[width=\columnwidth]{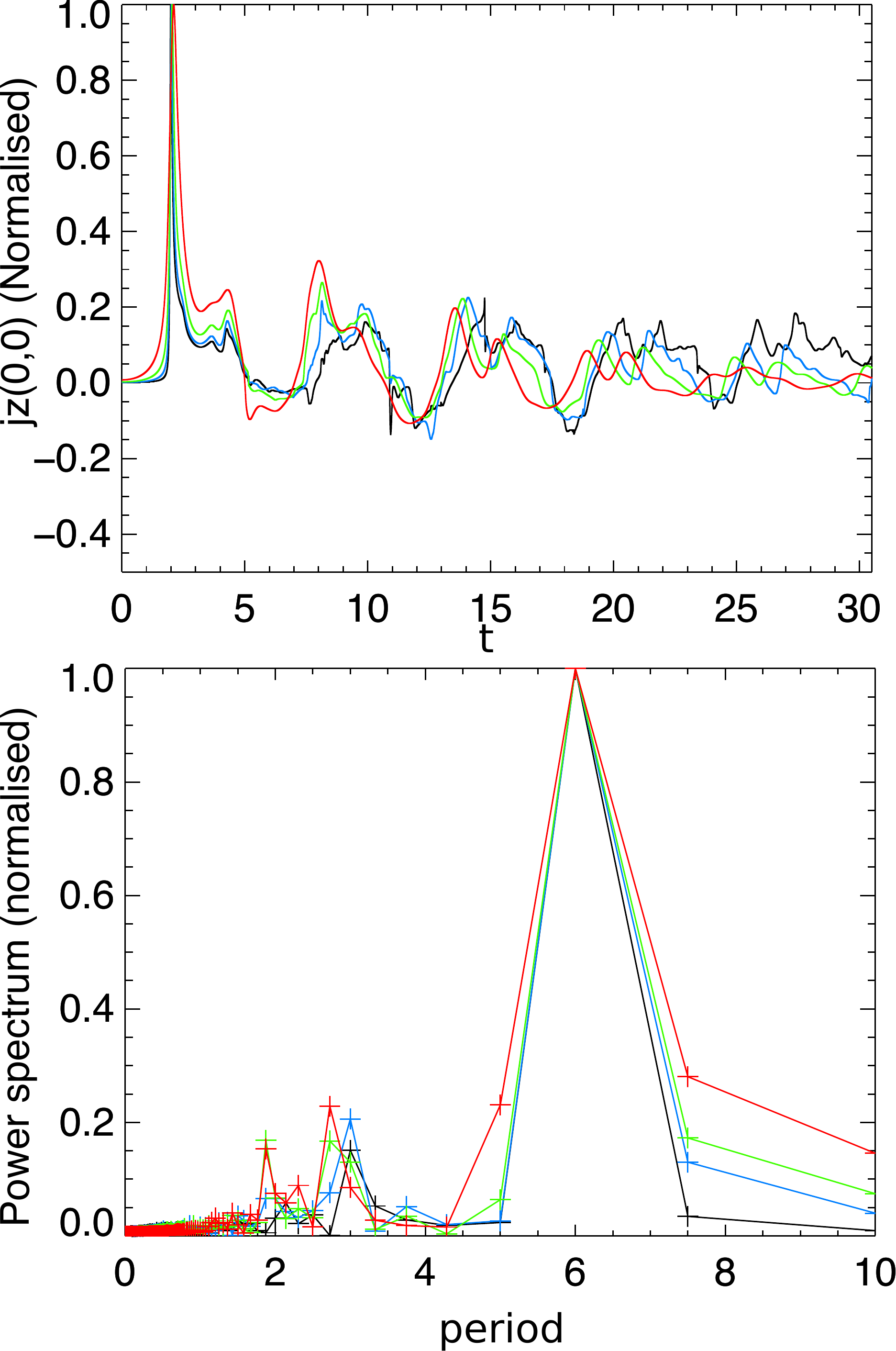}
      \caption{\textit{Top:} Curves of normalised $j_{z}$ recorded at the null for a range of resistivities with fixed perturbation amplitude  $j_{0}=0.1$ (sufficient for the initial  collapse to enter nonlinear evolution  in all cases shown).  
      Whilst it is clear that the signals are not identical and that additional  dissipation affects the curve (in particular, its smoothness), it is clear visually that the effect of resistivity upon the  period(s) is much weaker than in the linear case. 
      \textit{Bottom:} Normalised Fourier power spectra distribution.
      The dominant spectral bin,  is unchanged with $\eta$, indicating that any change in the main reversal period (change in current sheet sign and orientation) is only weakly dependent on resistivity, such that we cannot detect it with this sample. 
       In \textit{both panels}, red corresponds to the  lowest resistivity ($\eta=3\times10^{-5}$) and black to the highest ($\eta=1\times10^{-3}$).
       }
         \label{fig:NL_fixedamp}
   \end{figure}
   
\begin{table*}
\centering
\caption{Measured periodicities in the nonlinear simulations (see Section \ref{sec:NL} for explanation of different measurements).
	}
\label{tab:NL}
\begin{tabular}{ccccccc}
\hline
$\eta$     & $j_0$   & $t_{c}$ (measured) & $P$(FFT)  & $P$(reversal) & $l_{c}$ & $j_{c}$ \\
\hline
0.00003 & 0.1  & 1.99          & 5.46-6.67  & 5.69        & 0.28 & 192.55 \\
0.0001  & 0.1  & 2.01          & 5.46-6.67* & 5.60        & 0.28 & 87.38  \\
0.0003  & 0.1  & 2.04          & 5.46-6.67  & 5.52        & 0.26 & 37.03  \\
0.001   & 0.1  & 2.11          & 5.46-6.67  & 5.56        & 0.24 & 12.72  \\
\hline
0.0001  & 0.01 & 3.27          & 7.27-8.89  & 7.61        & 0.08 & 12.84  \\
0.0001  & 0.05 & 2.38          & 6.15-7.27  & 6.22        & 0.19 & 50.75  \\
0.0001  & 0.1  & 2.01          & 5.33-6.15* & 5.60        & 0.28 & 87.38  \\
0.0001  & 0.2  & 1.65          & 4.79-5.53  & 4.78        & 0.39 & 145.98\\
\hline
\end{tabular}
\end{table*}

   \begin{figure}
   \centering
   \includegraphics[width=\columnwidth]{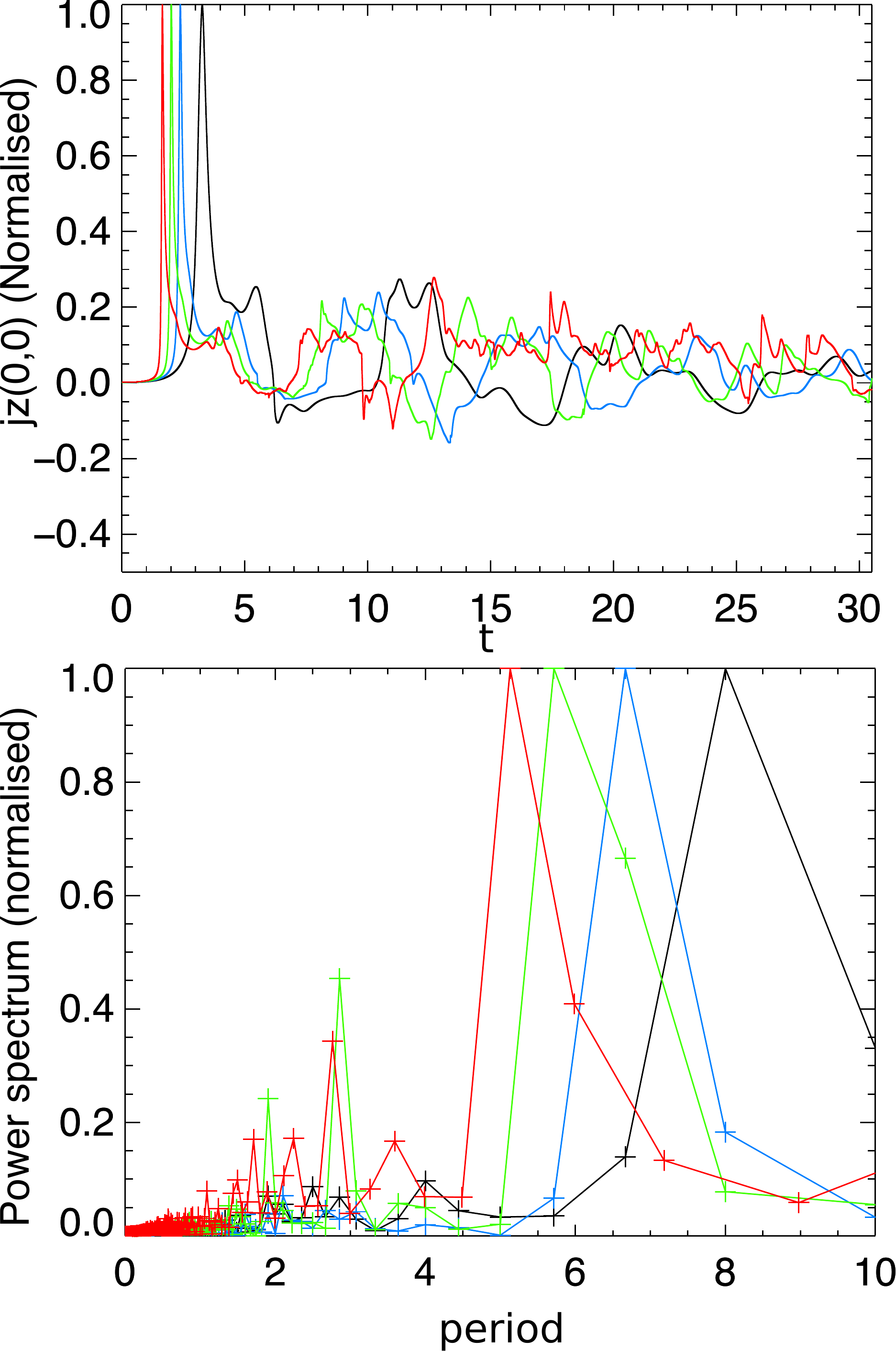}
      \caption{\textit{Top:} Curves of normalised $j_{z}$ recorded at the null for fixed resistivity $\eta=10^{-4}$ and perturbation amplitudes $j_{0}= 0.01$ (black), $0.05$ (blue), $0.1$ (green),$0.2$ (red).  
      It is visually clear that the larger amplitude perturbations form the initial current sheet more rapidly ($t_c$ occurs earlier, the initial maxima) and subsequently associated with higher frequency oscillations, both in terms of the reversal periods and the frequency of the secondary (same signed) peaks.
               \textit{Bottom:} Normalised Fourier power spectra distribution indicating the dominant period. There is a clear decrease in  the dominant period (associated with the main reversal cycle) as perturbation energy increases. Additionally, we note that as the amplitude is increased, spectral power in the lower-period (higher frequency) oscillations is also increased.
              }
         \label{fig:NL_varamp}
   \end{figure}

	We now consider the influence of increasing the amplitude of the perturbations in such closed systems beyond the threshold for the initial implosion to evolve nonlinearly {($j_{0}\gtrsim2.1\eta$ here)}. 
	As covered extensively in the literature \citep[e.g.][]{1979JPlPh..21..107F,1996ApJ...466..487M,Thurgood2018a,Thurgood2018b} and confirmed by experiment \citep[e.g.][]{1972JETPL..15...94S},  we anticipate the effect of exceeding this threshold to be the production during the initial implosion of a quasi-1D, highly-compressed and heated current sheet. 
	
	This can be seen in Figure \ref{fig:NLmov}, an embedded animation which shows the evolution of coloured contours of $j_{z}$, $\rho$ and $p$ for the case of $j_{0}=0.1$ and $\eta=10^{-4}$.
	By around $t=t_c=2.01$ such a current sheet has been formed and the initial implosion stalls.
	 Immediately after $t_c$, the inhomogeneity in the vicinity of the current sheet is obvious.
	  Shock structures form about the super-magnetosonic reconnection jets, namely  slow shocks on the flanks, and a fast termination shock at the head, which due to both magnetic- and gas-pressure gradients is driven in along the current sheet's length-wise axis, shortening it and choking off the reconnection flow.
	  The formation and properties of such shocks in the immediate aftermath of the initial implosion are described comprehensively in \citet{Thurgood2018b} and so we do not repeat the analysis here ---  instead we follow the longer-term evolution.  
	  
	  For the subsequent oscillations of the field and the current sheet (which evidentially occur in the later-time evolution of the animation of Figure \ref{fig:NLmov}), the plasma inhomogeneity that is generated has a dramatic effect. 
	  With regards to establishing a global oscillation of the field analogous to  the linear case, we note that the communication times from boundary to null point / current sheet are altered due to the mass-inhomogeneity  and localised heating (i.e. variable pressure field), altering both the Alfv\'en and sound speeds (hence, altering the fast and slow speeds). 
	  In addition to affecting the global communication times, the now-appreciable sound speed near the null will  allow for waves to traverse the null point  acoustically (possibly reflecting off the {density gradients at the} current sheet boundaries).
	  Furthermore, the current sheet and reconnection flow does not constitute a locally force-balanced state, and so `localised restoring forces' will act to alter the configuration, independently of global or standing modes of oscillation. 
	   For instance, at the ends of the current sheets (at the interface of the termination shock) both magnetic and gas pressure gradients conspire to choke off the outflow and drive in against the current sheet. This reduces its length and acts to promote current sheet reversal by a localised process of secondary collapse.
	Whilst in these particular simulations there will be contributions from these local force imbalances and reflections, we note that the local force imbalance means that OR could occur in the absence of reflections of outgoing waves at boundaries and the establishment of `standing' oscillations of the field (e.g. such a reversal cycle occurs in the case of \citealt{Thurgood2017}, which is an entirely open system). 
	Disentangling these various effects in a complicated nonlinear system is beyond the scope of this paper. 
	Instead, here we aim: (i) to highlight that different regimes are possible, and (ii) to begin exploring the resulting periodicity experimentally (via simulation).

	Figure \ref{fig:NL_fixedamp} (top panel) shows the (normalised) signal of $j_{z}$ measured at the null point for the simulations with fixed perturbation amplitude $j_{0}=0.1$ and resistivity in the range $\eta=3\times10^{-5}-1\times10^{-3}$. 
	It is immediately obvious from visual inspection that such signals are not as regularly sinusoidal as the linear case, instead exhibiting a quasi-periodicity characterised by a bursty signal containing double same-signed peaks, together with sign changes associated with current sheet reversals (cf. the signal in $j_{z}$ for $\eta=10^{-4}$ to the animation of Figure \ref{fig:NLmov}).
	Additionally, both the reversal period and the periods of multiple-extrema features are significantly shorter than the corresponding linear period for each equivalent value of $\eta$ (compare with Figure \ref{fig:linearsignals}). 
	Comparing amongst the curves in Figure \ref{fig:NL_fixedamp}, increasing resistivity appears to lead to a comparatively weak decrease in the main reversal period (relative to the  linear case), and an increasingly smooth time-variation.

	The corresponding power spectrum for the fixed-amplitude nonlinear runs 	is shown in Figure \ref{fig:NL_fixedamp} (bottom panel). 
	We find that for this set of runs there is no change in which  periodicity bin contains maximum power, and so we cannot quantitatively detect any apparent change in period with resistivity (within our spectral resolution). 
	However, as per our visual assessment, if we directly measure the main reversal period by simply determining the elapsed time between sign changes in $j_{z}$ over the first few cycles we do find some evidence that there may be a weak { inverse} scaling of the period with $\eta$. These direct measurements also confirm that the dominant spectral bin corresponds to the `main' periodicity associated with the current sheet reversals.
	 These measures (namely, the dominant spectral power bin and the directly measured reversal period) are summarised in Table \ref{tab:NL}.
	  Some secondary spectral peaks at lower period are also present, which we hypothesise are associated with the higher-frequency bursts in the signal, but we do not focus on them in this paper.
		 Thus, the dominant periodicity of nonlinear OR appears to have (at most) a weak { inverse} scaling with increasing $\eta$.  
	 This is a stark contrast with the case of linear OR where it is the only variable that matters. It is also {perhaps surprising, since} for our initially low value of $\beta$, that resistivity does effect a number of properties of the initial current sheet formed at the time of the implosion halting \citep{Thurgood2018a,Thurgood2018b}. 

	We also considered series of simulations with fixed resistivity ($\eta=10^{-4}$) and variable perturbation amplitude (all sufficiently large to permit nonlinear collapse). The signals $j_{z}(\mathbf{0},t)$ and their Fourier transform are shown in Figure \ref{fig:NL_varamp}, while key measurements  of these runs are  summarised in the lower section of Table \ref{tab:NL}.
	 In this case, we find that increasing the perturbation energy leads to higher-frequency oscillations, clearly visible in the different signals of $j_{z}(t)$. In the power spectrum, this shift to shorter periods is unambiguously detectable, with peaks separated into different periodicity bins.

Following \cite{2012A&A...548A..98M}, in Table \ref{tab:NL} we also record the maximum current density $j_{c}$ and current sheet length $l_{c}$ for both sets of nonlinear runs (both of which occur at the critical time $t_{c}$).
Decreasing the resistivity whilst maintaining a constant perturbation energy increases the current density, whilst the length of the current sheet is not strongly affected by changes in resistivity.
However, for fixed resistivity and increasing amplitude, we find that both the current density and the sheet length increase. 
Both aspects are well established: the increase in peak current density derives from the collapse proceeding to smaller scales across the current layer, while the current sheet length is predominantly determined by the radius at which nonlinear steepening becomes appreciable, when the cylindrical implosion becomes quasi-planar \citep[see e.g.][]{1996ApJ...466..487M,Thurgood2018b}. Thus larger perturbation energies result in longer current sheets.
From the measured value of $l_{c}$, we see therefore that $l_{c}$ appears to have an inverse scaling with the resulting periodicity. In other words, longer current sheets tend to be associated with shorter periods (faster reversals).
In the fixed resistivity case where we see weak (at best) scaling of the period with $\eta$, we see similarly small changes in recorded current sheet lengths.

\section{Dimensional considerations}\label{sec:dimensions}


\subsection{Energy threshold for nonlinearity}

The aforementioned requirement for nonlinearity to occur before a null point collapse is limited by resistivity is for our set-up $j_{0} \gtrsim 2.1 \eta$ (nondimensional). 
In Appendix \ref{app:threshold} we also derive the same condition beginning with the dimensional form of the background field, perturbed field, and magnetic resistivity. In doing so, it becomes clear that the limit is equivalent to 
\begin{equation}\label{eqn:dimthresh}
\frac{{\delta}E}{U_{0}(r=r_{c})}>1 \quad.
\end{equation}
This is a ratio of the total perturbation energy\footnote{
Here when we talk about `total energy', it can be considered as a column density or a total energy per unit length in the invariant $z$-direction, as we perform the energy integration only over the area in the $xz$-plane (i.e. as the problem is 2.5D).}
 (${\delta}E$) to the total potential energy of the background field evaluated  within the  linear diffusion radius $U_{0}(r=r_{c})$. If this inequality is not satisfied, the linear collapse should be diffusion limited before the perturbation overwhelms the background field.  Otherwise, it will begin to evolve nonlinearly.
 
 For our set up, these quantities are found to be:
\begin{eqnarray}
\delta E &=& \frac{32}{45}\cos^2(\alpha) \frac{b^2 l^2}{\mu_{0}}  \quad \mathrm{[J\,m^{-1}]} \\
U_{0}\left(r\right) &=& \frac{\pi}{4}\, \frac{b^2 r^4}{\mu_{0} l^2} \quad \mathrm{[J\,m^{-1}]}  \label{eqn:u0} \\
r_{c}^{2} &=& \left(\sqrt{\mu_{0}\rho_{0}}\,l\, b^{-1}\right) \eta \quad \mathrm{[m^2]}  \label{eqn:rc}
\end{eqnarray}
where $b$ (Tesla, $\mathrm{kg\,s^{-2}\,A^{-1}}$) corresponds to the strength of the field at the boundary, which is located at $l$ metres from the null, and it is understood that in this section of the paper $\rho_0$ and $\eta$ are dimensional quantities which correspond to a typical density ($\mathrm{ kg\,m^{-3}}$) and magnetic diffusivity ($\mathrm{ m^{2}\, s^{-1} }$). 
The perturbation amplitude relative to the background field strength on the boundary is here expressed in terms of the cosine of the separatrix angle $\alpha$  (which is related to the nondimensional number $j_{0}$ by $\cos\alpha=j_{0}/2$ ).

Combining Equations \ref{eqn:u0} and \ref{eqn:rc}, we find total energy associated with the background magnetic field in the linear diffusion region

\begin{equation}\label{eqn:u0rc}
U_{0}\left(r=r_{c}\right) = \frac{\pi}{4}\,\rho_{0}\eta^2 \quad \mathrm{[J\,m^{-1}]}.
\end{equation}
At this stage it becomes immediately clear that for typical astrophysical plasmas that essentially any perturbation of interest will violate equation \ref{eqn:dimthresh} and evolve nonlinearly.  
For example, if we take, say, $\rho_{0}\sim10^{-10}$ $\mathrm{kg\,m^{-3}}$ and $\eta\sim1$ $\mathrm{m^{2}\,s^{-1}}$ as reasonably representative of solar coronal plasma, then we  require  $\delta E \lesssim 10^{-10}$ $\mathrm{ J\,m^{-1}}$ for linear, resistively-limited collapse and OR. This is indicative that any collapse events energetic enough to observe will not be linear, resistively limited collapses. As such, any subsequent time-dependent reconnection will not be in the regime of linear OR as described in the initial models of e.g. \citet{1991ApJ...371L..41C}. Thus, the consideration of different regimes of OR such as the nonlinear regime presented in this paper is a necessary step forward.

\subsection{Simulated nonlinear periodicities under solar parameters}


It is not straightforward to unambiguously determine appropriate dimensional scales to assign  to models of linear null point fields (in the absence of further modelling context) as linear null points are scale-free. Instead we  consider the resulting time normalisation across a range of parameters that are consistent with  considering coronal densities in the range $10^{-12}<\rho_{0}<10^{-10}$ $\mathrm{kg\,m^{-3}}$ (noting that density has a relatively weak contribution as the square-root) and typical field strengths in the range $10^{-4}<b<10^{-1}$ $\mathrm{T}$ and length scales in the range $10^{5}<l<10^{7}$ $\mathrm{m}$.  In this case, the normalising time $t_{0}=l\sqrt{\mu_{0}\rho_{0}}b^{-1}\approx l\sqrt{10^{-6}\rho_{0}}b^{-1}$ (see Appendix \ref{appendixA}) will lie in the range $10^{-3}-1000$ seconds.

\section{Conclusions}\label{sec:discussion}
\subsection{Key Findings}
We have conducted numerical simulations of OR under a parameter study of resistivity and perturbation amplitudes in order to examine the effect on the resulting periodicity. 
Our key findings are

\begin{enumerate}

	\item We show, using a numerical simulation, that we can recover the main properties of the analytically-tractable systems of OR considered by \citet{1991ApJ...371L..41C,1992ApJ...393..385C,1992ApJ...399..159H,1993ApJ...405..207C}, which we call the {\it{linear OR}} regime.
	However, we find that once the perturbation energy sufficiently exceeds the threshold for nonlinear collapse (here, $j_{0}<\eta$ for linear collapse), more complex periodic signals are produced, replete with `bursty', same-signed peaks in current density and true reversals of the current sheet, which we call the {{nonlinear OR}} regime.
We find that the nonlinear OR regime is vastly different to that of the better-known linear OR cases.
	This threshold, which controls whether the initial implosion is itself linear, would likely be easily exceeded in astrophysical plasmas given typically  low Lundquist numbers. 

	\item In the nonlinear OR  case, we  find that for fixed perturbation energies the dependence of the measured spectra  on the resistivity (equivalently, Lundquist number) is sufficiently weak that it cannot be distinguished with our sampling period (our spectral resolution as per the sampling rate and time). 
	There is qualitative evidence that the dominant periodicity may have some weak positive scaling with the Lundquist number, but it is undoubtedly weaker than the linear OR  case where the change in period is clearly detectable in our simulations for the corresponding range of resistivity. 

	\item In the nonlinear OR regime, we find that by fixing the resistivity and then varying perturbation amplitude, the period is measurably affected by the amount of free energy available.
	 This is again in contrast to the linear OR case where the inverse Lundquist number (i.e. resistivity if we consider the length scales and maximum wave speed fixed) is the \textit{only} parameter that affects the period. 
	
\end{enumerate}

\subsection{Discussion}

{
With regards to key finding [1], and as per our discussion in Section \ref{sec:dimensions},  we note that the criteria for linear evolution is likely to be easily violated in astrophysical plasmas due to typically-low resistivity, and that perturbations that satisfy such a criteria are so energetically diminutive that they are inconsequential. }
%
%
%
	Further, as we have illustrated in Section \ref{sec:linear}, the period determined by the linear solution is essentially associated with a global oscillation of the field.
	In other words, it requires complete reflectivity at the boundaries, which may not be  appropriate in the solar atmosphere (see also  the discussion in \citealt{2007PhPl...14l2905L}).
	Whilst with partial reflection it may be the case that the linear period is not-much altered, it would also be inappropriate to apply the linear OR-type results on damping rates (disregarding the fact that the perturbation will likely be strong enough to violate the linearity condition  to begin with). 
Importantly, this appears not always to have been fully-appreciated in the literature, given that some previous studies of OR have mistakenly attempted to apply the periodicity and decay formulas of \citet{1991ApJ...371L..41C} to numerical simulations which are (apparently) reflection-free and clearly exhibit nonlinear behaviour.

	When the free energy contained within the initial disturbance is more substantial, we find the inevitable effect of the nonlinear implosion/collapse is that of a nonlinear scheme of OR.
	 In our simulations, both reflections (`global oscillations') and local dynamics around the highly-inhomogeneous and localised current sheet  contribute to setting the period.
	Remarkably, as per key finding [2] we find that these oscillations are not strongly dependent on plasma resistivity in the range considered.
	This is encouraging in that it suggests that it may be possible to make a direct quantitative comparison between simulated periodicities and those observed in the corona. While simulations must use much higher explicit resistivities than in reality, if the $\eta$-dependence of the period is weak as we have found, it may not be necessary to explicitly account for sub-grid dissipation or otherwise to extrapolate from scalings derived from accessible ranges of resistivity
	\footnote{
	We note however that it will be important to allow some avenue for controlled finite dissipation and consistent heating in the model  (ensuring energy conservation), and that current sheets should probably not be allowed to collapse to the grid scale. Here, all of our simulations have properly resolved current sheets/diffusion layers as detailed in Appendix \ref{app:boundary}. 
	}. 
	For example, the periodicity of current sheet reversals observed in  simulations more representative of the solar atmosphere, with model transition regions, dipole fields and external wave driving \citep[e.g. OR occurring in setups similar to][]{2017ApJ...837...94T} may be meaningful despite the requirement to use relatively high resistivity in such models (e.g. $S=10$ would be typical).
	
	Further, as per key finding [3], we find that instead the oscillations are affected by the energy available to the initial implosion. This sets the properties of the initial current sheet and the local inhomogeneity, indicating that such signals of OR could occur in open systems or in the absence of reflection (this is supported by \citealt{Thurgood2017} where one cycle of 3D OR was observed prior to a  guaranteed minimum reflection time, although the period was not considered quantitatively). 

\subsection{Towards OR for seismology}

Overall, our theoretical understanding of the range of periodicities produced in more realistic, nonlinear schemes of OR such as that considered here and elsewhere (see Introduction, \ref{sec:intro}) is currently insufficient to begin testing the possibility of OR underlying unexplained periodic phenomena in which reconnection has been implicated  (for example by determining whether the periods obtained are compatible with observed phenomena). 

In Section \ref{sec:dimensions}, it was shown that periods of $10^{-3}$ to $10^{3}$ seconds can be constructed for typical solar parameters, and \citet{McLaughlin2018} reported that QPPs have been detected with  periods ranging from a fraction of a second to several minutes. Thus, whilst promising, follow-up work (utilising the specific magnetic configuration one is comparing to) is needed to further constrain such a range of periodicities.

	One key result of this study is that perturbation amplitude (free energy) seems to have the strongest influence on the resulting periodicity --- 
	specifically, higher-energy collapses lead to measurably shorter periods and longer initial current sheets  (key finding [3]).
	We believe that this result should broadly generalise to other null-containing geometries  and setups given that it is easy to access the nonlinear phase of collapse during the initial implosion (key finding [1]).
	This finding is reminiscent of \cite{2012A&A...548A..98M}, who considered simulations of OR in a different system -- specifically one where OR is triggered by quasi-planar shocks converging on the null (which, after they pass through and then propagate away, leave behind a current sheet which oscillates).
		 They also found that the period was dependent on the perturbation amplitude (with shorter periods for higher amplitudes) and that the larger amplitude / shorter period systems were associated with longer initial current sheets. 
	
	\citet{2012A&A...548A..98M} hypothesised that  longer current sheets should  correspond to greater restoring forces participating in the reversal process (as they represent a greater disturbance to the background field).
	 They proposed that such larger restoring forces might explain the resulting higher-frequency oscillations by facilitating a more rapid reversal of the sheet.
	 	 However, for this to be the case the increase in such forces would have to be sufficient to compensate for the necessity to do work against the longer, hotter, and more massive current sheets which are produced by increasingly energetic initial implosions. 
	 	 	This interpretation would also be consistent with our finding that the nonlinear period does not strongly depend on resistivity (key finding [2]), as although the resistivity does affect a number of initial current sheet properties, it is known not to significantly influence the initial current sheet length in nonlinear null collapse (see \citealt{1996ApJ...466..487M} and \citealt{Thurgood2018a} for 2D and 3D, respectively). 	
	However, it is nontrivial to quantify what exactly constitutes and meaningfully quantifies the `net restoring force' in these nonlinear, inhomogeneous, and dissipative systems, and we have not yet been able to do so in a satisfactory manner. 
	We did consider a crude measure -- namely calculating the inwards force acting at the edge of the current sheet along its axis at $t_{c}$ -- and found that its magnitude is greater for longer current sheets. Furthermore, the increase is nonlinear (i.e. non-Hookean in the sense that the ratio of the force measured to the displacement, taken as $l_{c}$, is not constant but rather increases) which could be interpreted as consistent with the notion that this may ultimately drive the more rapid reversals.
	However, at this stage we are not confident that this is an appropriate measure from which to draw such a conclusion, hence why we did not present such data in Section \ref{sec:NL}, and so these particular comments should be read as preliminary. 
	 
		The influence of a number of other plasma variables and physical effects on the periodicity of OR remains to be studied, for example 
	more appreciable initial plasma pressures, 	
	plasma viscosity,
	guide-field and 3D effects (e.g. different 3D field line geometries),
	modified equations of state,
	 and thermal conductivity. 
	   All of the aforementioned parameters are expected to contribute to setting the scale at which the initial implosion is limited, and so control the initial current sheet properties. They will also control the plasma state about which subsequent oscillations occur, thus all have scope to affect the resulting periodicity.
	Additionally, the influence of reflectivity and boundary conditions needs to be further considered, in particular, the influence of partial-or-complete transmission of outgoing waves. 			
	Clearly, as per \citet{Thurgood2017}, current sheet reversals can be achieved in open systems (in the absence of reflected waves), although the quantitative aspects of such periods and their precise relation to easier-to-implement (and computationally efficient) closed-boundary models remain largely unexplored.

\begin{acknowledgements}
The authors acknowledge generous support from the  Leverhulme Trust and this work was funded by a Leverhulme Trust Research Project Grant: RPG-2015-075. The authors acknowledge IDL support provided by STFC. The computational work for this paper was carried out on HPC facilities provided by the Faculty of Engineering and Environment, Northumbria University, UK. JAM acknowledges STFC for support via ST/L006243/1. DIP acknowledges STFC for support via ST/N000714/1.
\end{acknowledgements}

%
\bibliographystyle{aa} 
\bibliography{references.bib} 

\begin{thebibliography}{43}
\expandafter\ifx\csname natexlab\endcsname\relax\def\natexlab#1{#1}\fi

\bibitem[{{Arber} {et~al.}(2016){Arber}, {Brady}, \&
  {Shelyag}}]{2016ApJ...817...94A}
{Arber}, T.~D., {Brady}, C.~S., \& {Shelyag}, S. 2016, \apj, 817, 94

\bibitem[{{Arber} {et~al.}(2001){Arber}, {Longbottom}, {Gerrard}, \&
  {Milne}}]{2001JCoPh.171..151A}
{Arber}, T.~D., {Longbottom}, A.~W., {Gerrard}, C.~L., \& {Milne}, A.~M. 2001,
  Journal of Computational Physics, 171, 151

\bibitem[{{Caramana} {et~al.}(1998){Caramana}, {Shashkov}, \&
  {Whalen}}]{1998JCoPh.144...70C}
{Caramana}, E.~J., {Shashkov}, M.~J., \& {Whalen}, P.~P. 1998, Journal of
  Computational Physics, 144, 70

\bibitem[{{Comisso} {et~al.}(2016){Comisso}, {Lingam}, {Huang}, \&
  {Bhattacharjee}}]{2016PhPl...23j0702C}
{Comisso}, L., {Lingam}, M., {Huang}, Y.-M., \& {Bhattacharjee}, A. 2016,
  Physics of Plasmas, 23, 100702

\bibitem[{{Craig} \& {Watson}(1992)}]{1992ApJ...393..385C}
{Craig}, I.~J. \& {Watson}, P.~G. 1992, \apj, 393, 385

\bibitem[{{Craig} \& {McClymont}(1991)}]{1991ApJ...371L..41C}
{Craig}, I.~J.~D. \& {McClymont}, A.~N. 1991, \apjl, 371, L41

\bibitem[{{Craig} \& {McClymont}(1993)}]{1993ApJ...405..207C}
{Craig}, I.~J.~D. \& {McClymont}, A.~N. 1993, \apj, 405, 207

\bibitem[{{Doyle} {et~al.}(2018){Doyle}, {Shetye}, {Antonova}, {Kolotkov},
  {Srivastava}, {Stangalini}, {Gupta}, {Avramova}, \&
  {Mathioudakis}}]{2018MNRAS.475.2842D}
{Doyle}, J.~G., {Shetye}, J., {Antonova}, A.~E., {et~al.} 2018, \mnras, 475,
  2842

\bibitem[{Dungey(1953)}]{dungey53}
Dungey, J. 1953, The London, Edinburgh, and Dublin Philosophical Magazine and
  Journal of Science, 44, 725

\bibitem[{Farrashkhalvat \& Miles(2003)}]{farrashkhalvat2003basic}
Farrashkhalvat, M. \& Miles, J. 2003, Basic Structured Grid Generation: With an
  Introduction to Unstructured Grid Generation, Referex Engineering
  (Butterworth-Heinemann Limited)

\bibitem[{{Forbes} \& {Speiser}(1979)}]{1979JPlPh..21..107F}
{Forbes}, T.~G. \& {Speiser}, T.~W. 1979, Journal of Plasma Physics, 21, 107

\bibitem[{{Hassam}(1992)}]{1992ApJ...399..159H}
{Hassam}, A.~B. 1992, \apj, 399, 159

\bibitem[{{Kupriyanova} {et~al.}(2016){Kupriyanova}, {Kashapova}, {Reid}, \&
  {Myagkova}}]{2016SoPh..291.3427K}
{Kupriyanova}, E.~G., {Kashapova}, L.~K., {Reid}, H.~A.~S., \& {Myagkova},
  I.~N. 2016, \solphys, 291, 3427

\bibitem[{{Kuznetsov} {et~al.}(2016){Kuznetsov}, {Zimovets}, {Morgachev}, \&
  {Struminsky}}]{2016SoPh..291.3385K}
{Kuznetsov}, S.~A., {Zimovets}, I.~V., {Morgachev}, A.~S., \& {Struminsky},
  A.~B. 2016, \solphys, 291, 3385

\bibitem[{{Longcope} \& {Priest}(2007)}]{2007PhPl...14l2905L}
{Longcope}, D.~W. \& {Priest}, E.~R. 2007, Physics of Plasmas, 14, 122905

\bibitem[{{Loureiro} {et~al.}(2007){Loureiro}, {Schekochihin}, \&
  {Cowley}}]{2007PhPl...14j0703L}
{Loureiro}, N.~F., {Schekochihin}, A.~A., \& {Cowley}, S.~C. 2007, Physics of
  Plasmas, 14, 100703

\bibitem[{{McClymont} \& {Craig}(1996)}]{1996ApJ...466..487M}
{McClymont}, A.~N. \& {Craig}, I.~J.~D. 1996, \apj, 466, 487

\bibitem[{{McLaughlin} {et~al.}(2009){McLaughlin}, {De Moortel}, {Hood}, \&
  {Brady}}]{2009A&A...493..227M}
{McLaughlin}, J.~A., {De Moortel}, I., {Hood}, A.~W., \& {Brady}, C.~S. 2009,
  \aap, 493, 227

\bibitem[{{McLaughlin} {et~al.}(2011){McLaughlin}, {Hood}, \& {de
  Moortel}}]{2011SSRv..158..205M}
{McLaughlin}, J.~A., {Hood}, A.~W., \& {de Moortel}, I. 2011, \ssr, 158, 205

\bibitem[{McLaughlin {et~al.}(2018)McLaughlin, Nakariakov, Dominique,
  Jel{\'i}nek, \& Takasao}]{McLaughlin2018}
McLaughlin, J.~A., Nakariakov, V.~M., Dominique, M., Jel{\'i}nek, P., \&
  Takasao, S. 2018, Space Science Reviews, 214, 45

\bibitem[{{McLaughlin} {et~al.}(2012{\natexlab{a}}){McLaughlin}, {Thurgood}, \&
  {MacTaggart}}]{2012A&A...548A..98M}
{McLaughlin}, J.~A., {Thurgood}, J.~O., \& {MacTaggart}, D. 2012{\natexlab{a}},
  \aap, 548, A98

\bibitem[{{McLaughlin} {et~al.}(2012{\natexlab{b}}){McLaughlin}, {Verth},
  {Fedun}, \& {Erd{\'e}lyi}}]{2012ApJ...749...30M}
{McLaughlin}, J.~A., {Verth}, G., {Fedun}, V., \& {Erd{\'e}lyi}, R.
  2012{\natexlab{b}}, \apj, 749, 30

\bibitem[{{Murray} {et~al.}(2009){Murray}, {van Driel-Gesztelyi}, \&
  {Baker}}]{2009A&A...494..329M}
{Murray}, M.~J., {van Driel-Gesztelyi}, L., \& {Baker}, D. 2009, \aap, 494, 329

\bibitem[{{Nakariakov} {et~al.}(2018){Nakariakov}, {Anfinogentov},
  {Storozhenko}, {Kurochkin}, {Bogod}, {Sharykin}, \&
  {Kaltman}}]{2018ApJ...859..154N}
{Nakariakov}, V.~M., {Anfinogentov}, S., {Storozhenko}, A.~A., {et~al.} 2018,
  \apj, 859, 154

\bibitem[{{Pontin}(2012)}]{Pontin3169}
{Pontin}, D.~I. 2012, Philosophical Transactions of the Royal Society of London
  A: Mathematical, Physical and Engineering Sciences, 370, 3169

\bibitem[{Priest(2014)}]{priest_book}
Priest, E. 2014, Magnetohydrodynamics of the Sun (Cambridge University Press)

\bibitem[{{Priest} \& {Forbes}(2000)}]{2000mare.book.....P}
{Priest}, E. \& {Forbes}, T. 2000, {Magnetic Reconnection}, 612

\bibitem[{{Priest} \& {Pontin}(2009)}]{2009PhPl...16l2101P}
{Priest}, E.~R. \& {Pontin}, D.~I. 2009, Physics of Plasmas, 16, 122101

\bibitem[{{Pugh} {et~al.}(2017){Pugh}, {Nakariakov}, {Broomhall}, {Bogomolov},
  \& {Myagkova}}]{2017A&A...608A.101P}
{Pugh}, C.~E., {Nakariakov}, V.~M., {Broomhall}, A.-M., {Bogomolov}, A.~V., \&
  {Myagkova}, I.~N. 2017, \aap, 608, A101

\bibitem[{{Roberts}(1971)}]{1971LNP.....8..171R}
{Roberts}, G.~O. 1971, in Lecture Notes in Physics, Berlin Springer Verlag,
  Vol.~8, Numerical Methods in Fluid DynamicsNumerical Methods in Fluid
  Dynamics, ed. M.~{Holt}, 171--177

\bibitem[{{Shen} {et~al.}(2018){Shen}, {Liu}, {Song}, \&
  {Tian}}]{2018ApJ...853....1S}
{Shen}, Y., {Liu}, Y., {Song}, T., \& {Tian}, Z. 2018, \apj, 853, 1

\bibitem[{{Syrovatskii}(1981)}]{syrovatskii1981review}
{Syrovatskii}, S.~I. 1981, \araa, 19, 163

\bibitem[{{Syrovatskii} {et~al.}(1972){Syrovatskii}, {Frank}, \&
  {Khodzhaev}}]{1972JETPL..15...94S}
{Syrovatskii}, S.~I., {Frank}, A.~G., \& {Khodzhaev}, A.~Z. 1972, Soviet
  Journal of Experimental and Theoretical Physics Letters, 15, 94

\bibitem[{{Tarr} {et~al.}(2017){Tarr}, {Linton}, \&
  {Leake}}]{2017ApJ...837...94T}
{Tarr}, L.~A., {Linton}, M., \& {Leake}, J. 2017, \apj, 837, 94

\bibitem[{{Threlfall} {et~al.}(2012){Threlfall}, {Parnell}, {De Moortel},
  {McClements}, \& {Arber}}]{2012A&A...544A..24T}
{Threlfall}, J., {Parnell}, C.~E., {De Moortel}, I., {McClements}, K.~G., \&
  {Arber}, T.~D. 2012, \aap, 544, A24

\bibitem[{{Thurgood} \& {McLaughlin}(2012)}]{2012A&A...545A...9T}
{Thurgood}, J.~O. \& {McLaughlin}, J.~A. 2012, \aap, 545, A9

\bibitem[{{Thurgood} \& {McLaughlin}(2013)}]{2013A&A...558A.127T}
{Thurgood}, J.~O. \& {McLaughlin}, J.~A. 2013, \aap, 558, A127

\bibitem[{{Thurgood} {et~al.}(2017){Thurgood}, {Pontin}, \&
  {McLaughlin}}]{Thurgood2017}
{Thurgood}, J.~O., {Pontin}, D.~I., \& {McLaughlin}, J.~A. 2017, \apj, 844, 2

\bibitem[{{Thurgood} {et~al.}(2018{\natexlab{a}}){Thurgood}, {Pontin}, \&
  {McLaughlin}}]{Thurgood2018a}
{Thurgood}, J.~O., {Pontin}, D.~I., \& {McLaughlin}, J.~A. 2018{\natexlab{a}},
  \apj, 855, 50

\bibitem[{{Thurgood} {et~al.}(2018{\natexlab{b}}){Thurgood}, {Pontin}, \&
  {McLaughlin}}]{Thurgood2018b}
{Thurgood}, J.~O., {Pontin}, D.~I., \& {McLaughlin}, J.~A. 2018{\natexlab{b}},
  Physics of Plasmas, 25, 072105

\bibitem[{{Van Doorsselaere} {et~al.}(2016){Van Doorsselaere}, {Kupriyanova},
  \& {Yuan}}]{2016SoPh..291.3143V}
{Van Doorsselaere}, T., {Kupriyanova}, E.~G., \& {Yuan}, D. 2016, \solphys,
  291, 3143

\bibitem[{{Wyper} {et~al.}(2017){Wyper}, {Antiochos}, \&
  {DeVore}}]{2017Natur.544..452W}
{Wyper}, P.~F., {Antiochos}, S.~K., \& {DeVore}, C.~R. 2017, \nat, 544, 452

\bibitem[{{Yamada} {et~al.}(2010){Yamada}, {Kulsrud}, \&
  {Ji}}]{2010RvMP...82..603Y}
{Yamada}, M., {Kulsrud}, R., \& {Ji}, H. 2010, Reviews of Modern Physics, 82,
  603

\end{thebibliography}
%

\begin{appendix}
\section{Nondimensionalisation and the LareXd code}\label{appendixA}
{
Following the details  in the LareXd user manual,  the normalisation is through the choice of three basic normalising constants, specifically:
\begin{eqnarray*}
x&=&L_0 \hat{x}\\
\mathbf{B}&=&B_0\hat{\mathbf{B}} \\
\rho&=&\rho_0 \hat{\rho}
\end{eqnarray*}
where quantities with and without a hat symbol are dimensional and nondimensional, respectively. We note that here subscript 0 refers to the normalisation constant and should not be confused with the use in Section \ref{sec:setup} to indicate an equilibrium or background quantity. These  are then used to define the normalisation of quantities with derived units through
\begin{eqnarray*}
v_{0}&=&\frac{B_{0}}{\sqrt{\mu_{0}\rho_{0}}}\\
P_{0}&=&\frac{B^{2}_{0}}{\mu_{0}} \\
t_0&=&\frac{L_0}{v_0}\\
j_{0}&=&\frac{B_{0}}{\mu_{0}L_{0}}\\
E_0&=&v_0 B_0\\
\varepsilon_0&=&v_0^2
\end{eqnarray*}
so that $\mathbf{v}=v_0\hat{\mathbf{v}}$, $\mathbf{j}=j_0\hat{\mathbf{j}}$, $t=t_0\hat{t}$ and 
$P=P_0\hat{P}$ etc. 
Applying this normalisation to the ideal MHD equations simply removes the vacuum 
permeability $\mu_0$. In resistive MHD, this scheme leads naturally to a resistivity normalisation:
 \begin{displaymath}
 \hat{\eta}=\frac{\eta}{\mu_0 L_0 v_0}
\end{displaymath}
or $\eta_0=\mu_0 L_0 v_0$. Since $v_0$ is the normalised Alfv\'en speed this means that 
$\hat{\eta}=1/S$ where $S$ is the Lundquist number as defined by the basic normalisation constants. 
}

The simulation is the numerical solution of the nondimensional, resistive MHD equations: ({NB: we drop the hat from this point onwards in the appendix, and throughout the main paper all quantities are nondimensional)}
\begin{eqnarray}
\frac{\mathrm{D}\rho}{\mathrm{D}t}&=&-\rho \nabla\cdot \mathbf{v}\\
\frac{\mathrm{D}\mathbf{v}}{Dt}&=&\frac{1}{\rho}(\nabla\times\mathbf{B})\times\mathbf{B}
-\frac{1}{\rho}\nabla p + \mathbf{F}_{shock}\\
\frac{\mathrm{D}\mathbf{B}}{\mathrm{D}t}&=&(\mathbf{B}\cdot\nabla)\mathbf{v}-\mathbf{B}
(\nabla\cdot\mathbf{v})-\nabla\times(\eta\nabla\times\mathbf{B})\\
\frac{\mathrm{D}\varepsilon}{\mathrm{D}t}&=&-\frac{p}{\rho}\nabla\cdot\mathbf{v}+\frac
{\eta}{\rho}j^{2} + \frac{\mathbf{H}_{visc}}{\rho}\\
\mathbf{j} &=& \mathbf{\nabla}\times\mathbf{B}\\
\mathbf{E} &=& -\mathbf{v}\times\mathbf{B}+\eta\mathbf{j}\\
p &=& \varepsilon\rho\left(\gamma-1\right)
\end{eqnarray}
which are solved on a Cartesian grid using the 2D version of the code (where ${\partial}/{\partial}z=0$ is hard-coded). All results presented are in non-dimensional units. Algorithmically, the code solves the ideal MHD equations explicitly using a Lagrangian remap approach and includes the resistive terms using explicit subcycling \citep{2001JCoPh.171..151A,2016ApJ...817...94A}. The solution is fully nonlinear and captures shocks via an edge-centred artificial viscosity approach \citep{1998JCoPh.144...70C}, where shock viscosity is applied to the momentum equation through $\mathbf{F}_{shock}$ and heats the system through $\mathbf{H}_{visc}$. Extended MHD options available within the code, such as the inclusion of Hall terms, were not used in these simulations. Full details of the code can be found in the original paper \citep{2001JCoPh.171..151A} and the users' manual.

\section{Boundary conditions and grid setup}\label{app:boundary}

In all simulations presented here we solve for the Cartesian domain $|x,y|\le1$. In practice we exploit the symmetry of the problem to only compute a solution in the quarter-plane $(x,y)\le1$, and so apply the appropriate symmetry / antisymmetry conditions on the `internal' computational boundaries ($x=0$ and $y=0$). At the `external' boundaries ($x=1$ and $y=1$) we permit  no flow through or along the boundary ($\mathbf{v}=\mathbf{0}$) with zero-gradient conditions taken on $\rho$ and $\varepsilon$, and take the magnetic field as line-tied (zero-gradient tangential and fixed normal components).
	The suitability of these boundary conditions, and overall stability of the setup, was checked by runs with and without perturbations  (in the null collapse case, recall that the force imbalance is localised to $r<1$). In these tests we found that there was no undesirable behaviour such as the launching of spurious waves from the outer boundary or erroneous current formation at the boundary, and that the state at the boundary remains static until the outwardly propagating part of a given perturbation reaches it. 
	The implementation and accuracy of the symmetry conditions were checked simply by re-running some simulations in the whole domain, and we find perfect agreement.

	To adequately resolve the small scale features produced by the collapse, especially in the lower resistivity cases, grid stretching is employed to concentrate resolution in the vicinity of the current sheets. The grids cell boundary positions $x_{b}$ along the $x$-direction  are distributed according the transformation \citep{1971LNP.....8..171R,farrashkhalvat2003basic}:
	\begin{equation}
	x_{b} = x_{\mathrm{max}} 
		\frac{\left(\lambda_x+1\right) - \left(\lambda_x-1\right) \left(\frac{\lambda_x+1}{\lambda_x-1} \right)^{1-\xi_x}
		}{\left(\frac{\lambda_x+1}{\lambda_x-1} \right)^{1-\xi_x} +1 }
	\end{equation}
	where $\xi_{x,i}$ is a uniformly-distributed computational coordinate $\xi_x\in[0,1]$ subdivided amongst the number of cells used in the $x$ direction. The degree of grid clustering at the $x=0$ is controlled by the stretching parameter $\lambda_x$. Likewise, the same form and parameters are used for the distribution of cells in $y$. 
	In our final runs we chose $\lambda_x = \lambda_Y = 1.01$, then performed simulations with increasing numbers of cells up to a maximum of $nx=ny=1024$,  (effectively, $2048^2$ cells given the symmetry). 
	Combined with the stretching, this yields a resolution of ${\Delta}x_{min}={\Delta}y_{min}\approx5.1\times10^{-5}$ and ${\Delta}x_{max}={\Delta}y_{min}\approx2.6\times10^{-3}$ in the finest regions of the grid. 
	   Each of these final simulations is in good agreement with a simulation at half the stated resolution (half of the cells in each dimension), in a qualitative sense during the evolution of the implosion and in the quantitative of producing minimal changes in measured quantities at $t_{c}$. Furthermore, we note that we have performed exhaustive testing of the requirements to resolve collapses in \citet{Thurgood2018b}, and found that the initial implosion obeys analytically predicted scaling laws. Such scaling results on the initial implosion carry-over to this paper and give us confidence / analytical verification that the  most challenging aspect of this problem, the initial implosion, is properly resolved. 
	   
	 
\section{Amplitude required for departure from linearity}\label{app:threshold}
We can re-write the background and perturbation fields (Equations \ref{eqn:b0} and \ref{eqn:b1}) in the dimensional form

\begin{eqnarray}
		\mathbf{B}_{0} &=& b \left[ \frac{y}{l}, \frac{x}{l}, 0\right]  \\
		\mathbf{B}_{1} &=& b\cos\left(\alpha\right)\left[
			\frac{y}{l}\left(\left(\frac{x}{l}\right)^2-1 \right),
		 	-\frac{x}{l}\left(\left(\frac{y}{l}\right)^2-1 \right), 
		 0\right]  
\end{eqnarray}
where $b$ is the field strength of $B_{0}$ at radius $r=l$ from the null point (origin) and is measured in Tesla, and $x,y,l$ are measurements of length in metres. The perturbation amplitude is expressed in terms of the separatrix angle at the null $\alpha$ and is related to the initial nondimensional current density $j_{0}$ by $j_{0} = 2\cos\alpha$.

The total energy introduced by the perturbing field may be found by integration of its energy density over the \textit{cartesian domain}
\begin{equation}
\delta E = \int \frac{B_1^{2}}{2\mu_{0}} \mathrm{dA}=
\frac{32}{45}\cos^2(\alpha) \frac{b^2 l^2}{\mu_{0}}  \quad \mathrm{[J\,m^{-1}]} .
\end{equation}
This energy is conserved within the imploding region until reaching the stage where diffusion becomes appreciable.
During the linear phase of collapse, it will begin to assume a cylindrical distribution to match the equilibrium Alfv\'en speed profile.

The total energy in the 2D plane associated with the field $B_{0}$ within a \textit{cylinder} of radius $r$ is
\begin{equation}
U_{0}\left(r\right) = \int \frac{B_0^{2}}{2\mu_{0}}\mathrm{dA} = \frac{\pi}{4}\, \frac{b^2 r^4}{\mu_{0} l^2} \quad \mathrm{[J\,m^{-1}]}  .
\end{equation}
Nonlinear evolution will begin to proceed once the perturbation reaches a sufficiently small radius that its own magnetic energy (and associated magnetic pressure/Lorentz force) becomes comparable to that of the background field. Thus, to enter a nonlinear phase of evolution we require that $\delta E$ exceeds $U_{0}(r)$ at some radius greater than the linear diffusion radius. In other words, we require
\begin{equation}
\frac{{\delta}E}{U_{0}(r=r_{c})}>1 \quad ,
\end{equation}
\textit{for the collapse and subsequent OR to be nonlinear}. Note that it might be more appropriate to consider the perturbation magnetic energy as  $\delta E/2$, since equipartition with kinetic energy during the ideal phase is quickly achieved. However, this does not materially change what is anyway an order-of-magnitude estimate of the threshold.

We can estimate the diffusion radius by requiring equality of the Alfv\'en speed at a given radius be equal to a diffusion speed based on a length scale corresponding to such a radius
\begin{align}
\frac{|B_{0}(r_{c})|}{\sqrt{\mu_{0}\rho_{0}}} &= \frac{\eta}{r_{c}} \nonumber \\
|B_{0}(r_{c})| &= \frac{br_{c}}{l} \nonumber \\
\longrightarrow r_{c}^{2} &= \left(\sqrt{\mu_{0}\rho_{0}} l b^{-1}\right) \eta \quad \mathrm{[m^2]}  \quad .
\end{align}

These equations constitute those discussed in Section \ref{sec:dimensions}. 

Combining the three equations in the form $\delta E > U_{0}(r=r_{c})$, making cancellations, dropping dimensional quantities, and expressing the amplitude in terms of initial nondimensional current density$j_{0}$, we recover the equivalent limit 
\begin{equation}
j_{0} \gtrsim \sqrt{\frac{45\pi}{32}}\eta \approx 2.1 \eta
\end{equation}
which was used in the numerical sections of the paper (Sections \ref{sec:linear} and \ref{sec:NL}).
\end{appendix}

\end{document}